\documentclass[12pt]{iopart}
\usepackage{iopams}
\usepackage{subfigure}
\usepackage{graphicx}

\begin{document}

\title[Hyperstars]{The influence of strength of hyperon-hyperon interactions on  neutron star properties.}

\author{I. Bednarek and R. Manka
\footnote[3]{manka@us.edu.pl}
}

\address{Department of Astrophysics and Cosmology, Institute of Physics,
 University of Silesia, Uniwersytecka 4, PL-40-007 Katowice, Poland}

\begin{abstract}
An equation of state of neutron star matter with  strange baryons
has been obtained. The effects of the strength of hyperon-hyperon
interactions on the equations of state constructed for the  chosen
parameter sets have been analyzed. Numerous neutron star models
show that the appearance of hyperons is connected with the
increasing density in neutron star interiors. The performed
calculations have indicated that the change of the hyperon-hyperon
coupling constants affects the chemical composition of a neutron
star. The obtained numerical hyperon star models  exclude large
population of strange baryons in the star interior.
\end{abstract}

\maketitle

\section{Introduction}
The analysis of the role of strangeness in nuclear structure in
the aspect of multi-strange system is of great importance for
relativistic heavy-ion collisions and for astrophysics for the
description of hyperon star matter. At the core of a neutron star
the matter density ranges from a few times the density of normal
nuclear matter to about an order of a magnitude higher. Thus
exotic forms of matter such as hyperons are expected to emerge in
the interior of a neutron star. The appearance of these additional
degrees of freedom and their impact on a neutron star structure
have been the subject of extensive studies \cite{glen},
\cite{bema}, \cite{weber}, \cite{gal}. Properties of matter at
such extreme densities are of particular importance in determining
forms of equations of state relevant to neutron stars and
successively in examining their global parameters.\\ The existence
of bound strange hadronic matter which in addition to nucleons
contains also hyperons has  profound consequences for
astrophysics. The starting point in studying the role of
strangeness in nuclear structure is the knowledge of the
properties of a single hypernucleus. Quantum chromodynamics should
be applied to the theoretical description of hadronic systems
owing to the fact that it constitutes the fundamental theory of
strong interactions.  However, at the hadronic energy scale where
the experimentally observed  degrees of freedom are not quarks but
hadrons  the direct description of nuclei in terms of QCD become
inadequate and thus alternative approaches had to be formulated.
One of them is quantum hadrodynamics (QHD) \cite{walecka},
\cite{walecka2} which gives quantitative description of the
nuclear many body problem. QHD is a relativistic quantum field
theory in which nuclear matter description in terms of baryons and
mesons is provided. The original model (QHD-I) contains nucleons
interacting through the exchange of simulating medium range
attraction $\sigma$ meson and $\omega$ meson responsible for short
range repulsion. Extension (QHD-II) of this theory
\cite{bog77},\cite{bodmer} includes also the isovector meson $\rho
$. Theoretical description of strange hadronic matter, which
satisfactorily reproduces nucleon-nucleon and hyperon-nucleon
data, has been given within the non-relativistic and relativistic
mean field models. This approach is based on the notion of the
meson-exchange model in which baryons interact through the
exchange of mesons. In addition to the $\sigma$, $\omega$ and
$\rho$ mesons these models contain $\sigma^{\ast}$ and $\phi$
mesons, introduced in order to
reproduce the strong attractive hyperon-hyperon interactions \cite{Schaffner}.\\
The vector coupling constants are chosen according to SU(6)
symmetry  whereas the scalar coupling constants are fixed to
hypernuclear data. Recent reports on observations of a
${}^{6}_{\Lambda\Lambda}He$ hypernucleus made by Takahashi et al.
\cite{Takahashi} provided information on the $\Lambda-\Lambda$
interaction energy $\delta B_{\Lambda\Lambda}=1.01\pm
0.20^{+0.18}_{-0.11}$ MeV. This allows one to determine the value
of the $\Lambda$ well depth in $\Lambda$ matter at density $0.5
\rho_0$ ($\rho_0$ denotes the saturation density) at the level of
$U_{\Lambda}^{\Lambda}\simeq 5$ MeV \cite{Song}. The results of earlier
experiments \cite{Prowse}, \cite{Dalitz}, \cite{Danysz},
\cite{Aoki}, \cite{Dover} give the value of
$U_{\Lambda}^{\Lambda}$ estimated with the use of Nijmegen model D
 at the level of $\simeq 20$ MeV
\cite{Schaffner}. This paper examines the implications of the
strength of hyperon-hyperon coupling constants on the
$\beta$-equilibrated hyperon star matter, hence doing  all
calculations within the framework of relativistic mean field model
with two parameterizations, namely the standard TM1 \cite{tm1} and
TMA \cite{Toki}.  The results have been obtained for two cases:
the weak and strong $Y-Y$ interactions. First, the properties of
the isospin symmetric strange hadronic matter have been
investigated. The obtained saturation curves resemble those
obtained by Song et al.\cite{Song}. The results for the two models
(TM1 and TMA) are very similar. The extension of the considered
model to the $\beta$ equilibrated asymmetric hyperon star model
has been done in the subsequent section. As a  result of this the
composition, the equation of state and the hyperon star structure
for the two parameterizations have been obtained.
\section{Hypernuclei}
Information on  single hypernuclei can be summarized in the
following points:
\begin{itemize}
\item $\Lambda$ hypernuclei: there is a large amount of data on
the binding energies $B_{\Lambda}$ of $\Lambda$'s bound in various
single particle orbitals in hypernuclei. This enables us  to study
deeply bound states inside the nucleus over an extensive range of
mass number. An analysis of these data with the use of
Skyrme-Hartree-Fock \cite{Millner} model gives the potential depth
of a single $\Lambda$ in nuclear matter at the value of
\begin{equation}
U^{(N)}_{\Lambda}\simeq 27-30 MeV
\end{equation}
which corresponds to 1/3-1/2 of the nucleon well depth
$U^{(N)}_{N}$ (in this text all potentials are considered as
attractive but the convention of positive sign has been used).
\item $\Sigma$ hypernuclei \cite{Mares}: the experimental status
of $\Sigma$ nucleus potential still remains controversial. The
calculations of $\Sigma$ hypernuclei have been based on analysis
of $\Sigma^{-}$ atomic data. Phenomenological analysis of level
shifts and widths in $\Sigma^{-}$ atoms made by Batty et al.
\cite{ksi} indicates that the $\Sigma$ potential is attractive
only at the nuclear surface changing into repulsive one in
increasing density. The small attractive component of this
potential is not sufficient to form bound $\Sigma$-hypernuclei.
Balberg et al. in their paper \cite{Balberg} show that the system
which includes $\Sigma$, $\Lambda$ hyperons and nucleons will be
unstable with respect to strong reactions
$\Sigma+N\rightarrow\Lambda+N$ (78 MeV),
$\Sigma+\Sigma\rightarrow\Lambda+\Lambda$ (156 MeV),
$\Sigma+\Lambda\rightarrow\Xi+N$ (50 MeV),
$\Sigma+\Xi\rightarrow\Lambda+\Xi$ (80 MeV) \cite{Stoks}, in
parenthesis the $Q$ values for each reaction are given. \item
$\Xi$ hypernuclei - for the $\Xi$ hypernuclei there exist a few
emulsion events reported in literature indicating  the existence
of a bound system. The interpreted data give the potential  of a
$\Xi$ in nuclear matter with a depth of
\begin{equation}
U_{\Xi}^{(N)}\simeq 20-25 \hspace{0.5cm}MeV.
\end{equation}
\item $\Lambda\Lambda$ hypernuclei: the properties of single
hypernuclei are one aspect of studying strangeness in nuclear
systems; the other  is connected with multi-strange systems. The
extrapolation to a multi-strange system is based on the data
concerning double $\Lambda$-hypernuclei.
 Data on $\Lambda\Lambda$
hypernuclei are extremely scarce. Observation of double-strange
hypernuclei $\Lambda\Lambda$ provide information about the
$\Lambda-\Lambda$ interaction. Several events have been identified
which indicate an attractive $\Lambda\Lambda$ interaction. The
analysis of the data allows one to estimate the binding energies
of ${}_{\Lambda\Lambda}^{6}He$, ${}_{\Lambda\Lambda}^{10}Be$ and
${}_{\Lambda\Lambda}^{13}B$. The measurement of the masses of
double-$\Lambda$ hypernuclei gives information on the sum of the
binding energy of the two $\Lambda$ hyperons $B_{\Lambda\Lambda}$
and the $\Lambda-\Lambda$ interaction energy $\Delta
B_{\Lambda\Lambda}$.  These two quantities can be defined as
\begin{eqnarray}
B_{\Lambda\Lambda}({}_{\Lambda\Lambda}^{A}Z)&=&B_{\Lambda}({}_{\Lambda\Lambda}^{A}Z)+B_{\Lambda}({}_{\Lambda}^{A-1}Z)\\
\nonumber \Delta
B_{\Lambda\Lambda}({}_{\Lambda\Lambda}^{A}Z)&=&B_{\Lambda}({}_{\Lambda\Lambda}^{A}Z)-B_{\Lambda}({}_{\Lambda}^{A-1}Z).
\end{eqnarray}
Table \ref{tab:lambda} compiles experimental values of the two
 observables mentioned above  \cite{Dalitz}, \cite{Aoki},
\cite{Danysz}, \cite{Prowse}.
\begin{table}
\begin{center}
\begin{tabular}{llll}\\
  \hline\hline
Hypernucleus&$B_{\Lambda\Lambda}[MeV]$&$\Delta B_{\Lambda\Lambda}$[MeV]\\
\hline ${}_{\Lambda\Lambda}^{6}He$&$10.9\pm 0.6$&$4.7\pm 0.6$\\
\hline${}_{\Lambda\Lambda}^{10}Be$&$17.7\pm 0.4$ &$4.3 \pm 0.4$ \\
\hline${}_{\Lambda\Lambda}^{13}B$& $27.5\pm 0.7$ & $4.8 \pm 0.7$\\
\hline\hline \label{tab:lambda}
\end{tabular}
\end{center}
\caption{The value of $\Delta B_{\Lambda\Lambda}$ and
$B_{\Lambda\Lambda}$ of the known double $\Lambda$-hypernuclei.}
\end{table}
The obtained values of $\Delta B_{\Lambda\Lambda}$ indicate that
the $\Lambda-\Lambda$ interaction is attractive and rather strong.
The value of $\Delta B_{NN}$ equals 6-7 MeV for comparison.
Following the estimation made by Schaffner et al.\cite{Schaffner}
it is possible to approximately determine the ratio of the
$\Lambda$ well depth in $\Lambda$ matter and nucleon well depth in
nuclear matter
\begin{equation}
\frac{U_{\Lambda}^{(\Lambda)}}{U_{N}^{(N)}}=\frac{V_{\Lambda\Lambda}}{V_{NN}}\frac{1/4}{3/8}\frac{1}{2}\label{ll}
\end{equation}
where $1/2$ stands for a nucleon and $\Lambda$ density ratio and
$(1/4)/(3/8)$ denotes spin-isospin weights of spatially symmetric
two-body configurations.
 Taking into account the value of $V_{\Lambda\Lambda}\equiv
\Delta B_{\Lambda\Lambda}\simeq 4-5$ MeV and the value
$V_{NN}\simeq 6-7$ MeV one can estimate the ratio
$V_{\Lambda\Lambda}/V_{NN}\approx 3/4$ and thus the equation
(\ref{ll}) gives
\begin{equation}
\frac{U_{\Lambda}^{(\Lambda)}}{U_{N}^{(N)}}\approx\frac{1}{4}.
\end{equation}
For $U_{N}^{N}\simeq 80$ MeV the estimated value of
$U_{\Lambda}^{\Lambda}$ potential is $\simeq 20$ MeV.\\
However, the recent data analysis of double hypernucleus
${}_{\Lambda \Lambda}^{6}He$ done by Takahashi et al.
\cite{Takahashi} gives the following value of the $\Lambda
-\Lambda$ interaction energy $\Delta B_{\Lambda \Lambda }=1.01\pm
0.20^{+0.18}_{-0.11}$ MeV what indicate that the interaction is
much weaker. The potential well depth evaluated for this data has
the value $U^{(\Lambda )}_{\Lambda }\simeq 5$ MeV.
\end{itemize}
\section{The model}
The theoretical description of the properties of strange hadronic
matter is given within the relativistic mean field approach. The
considered model involves baryons interacting through the exchange
of simulating medium range attraction $\sigma$ meson and $\omega$
meson responsible for short range repulsion. The model also
includes the isovector meson $\rho $. In order to reproduce
attractive hyperon-hyperon interaction two additional
hidden-strangeness mesons, which do not couple  to nucleons, have
been introduced, namely the scalar meson $f_0(975)$ (denoted as
$\sigma^{\ast}$) and the vector meson $\phi(1020)$.\\ The
effective Lagrangian function for the system can be written as a
sum of a baryonic part including the full octet of baryons
together with baryon-meson interaction terms and a mesonic part
\begin{equation} \label{lag}
{\mathcal{L}}={\mathcal{L}_{B}}+{\mathcal{L}_{M}}.
\end{equation}
The interacting baryons are described by the Lagrangian function
${\mathcal{L}_{B}}$ which is given by
\begin{equation}
{\mathcal{L}_{B}}=\sum_{B}
\bar{\psi}_B(i\gamma^{\mu}D_{\mu}-m_B+g_{\sigma
B}\sigma+g_{\sigma^{\ast}B}\sigma^{\ast})\psi_B.
\end{equation}
where  $B$ stands for $N, \Lambda, \Sigma, \Xi$ and  $\Psi_B^T
=(\psi_N,\psi_{\Lambda},\psi_{\Sigma},\psi_{\Xi})$.  The covariant
derivative $D_{\mu}$ is defined as
\begin{equation}
D_{\mu}=\partial_{\mu}+ig_{\omega B}\omega_{\mu}+ig_{\phi
B}\phi_{\mu}+ig_{\rho B}I_{3B}\tau^a\rho^a_{\mu}.
\end{equation}
The Lagrangian density for meson fields takes the form
\begin{eqnarray}
{\mathcal{L}_{M}} & = & \frac{1}{2}\partial _{\mu }\sigma \partial ^{\mu }\sigma -U(\sigma )+ \frac{1}{2}\partial _{\mu }\sigma^* \partial ^{\mu }\sigma^*- \frac{1}{2}m^{2}_{\sigma^*}\sigma^{*2}+ \nonumber \\
 & + &\frac{1}{2}m^{2}_{\phi}\phi_{\mu}\phi ^{\mu} -\frac{1}{4}\phi_{\mu \nu}\phi^{\mu \nu}
 -\frac{1}{4}\Omega _{\mu \nu }\Omega ^{\mu \nu }+\frac{1}{2}m_{\omega }^2\omega _{\mu }\omega ^{\mu }+ \nonumber \\
 & - & \frac{1}{4}R_{\mu \nu }^{a}R^{a\mu \nu }+\frac{1}{2}m_{\rho} ^2\rho^{a}_{\mu }\rho^{a\mu }+\frac{1}{4}c_{3}(\omega _{\mu }\omega ^{\mu })^{2}. \label{lag1}
\end{eqnarray}
The field tensors \( R_{\mu \nu }^{a} \), \( \Omega _{\mu \nu }
\), \( \phi_{\mu \nu } \) are defined as
\begin{equation}
R_{\mu \nu }^{a}=\partial _{\mu }\rho^{a}_{\nu }-\partial _{\nu
}\rho^{a}_{\mu }+g_{\rho }\varepsilon _{abc}\rho_{\mu
}^{b}\rho_{\nu }^{c},
\end{equation}
\begin{equation}
\Omega _{\mu \nu }=\partial _{\mu }\omega _{\nu }-\partial _{\nu
}\omega _{\mu },\hspace{2cm}\phi_{\mu \nu }=\partial _{\mu }\phi
_{\nu }-\partial _{\nu }\phi _{\mu }
\end{equation}
The potential function \( U(\sigma ) \) possesses a polynomial
form
\begin{equation}
U(\sigma )=\frac{1}{2}m^{2}_{\sigma}\sigma ^{2}+\frac{1}{3}g_{3}\sigma
^{3}+\frac{1}{4}g_{4}\sigma ^{4}.
\end{equation}
The baryon mass is denoted by $m_B$ whereas \( m_{M} \) $(M=
\sigma ,\omega ,\rho ,\sigma^*,\phi )$ are masses assigned to the
meson fields. The derived equations of motion constitute a set of
coupled equations which have been solved in the mean field
approximation. In this approximation meson fields are separated
into classical mean field values and quantum fluctuations which
are not included in the ground state
\begin{center}
\begin{tabular}{lll}\\
  \hline
$\sigma  = \overline{\sigma}$ +$ s_0$ &$ \sigma^*  =
\overline{\sigma}^*$
+ $s^*_0$ &   \\
$\phi_{\mu}  = \overline{\phi}_{\mu} + f_0\delta_{\mu 0}$ &
$\omega_{\mu} = \overline{\omega}_{\mu}+ w_{0}\delta_{\mu 0}$ &
$\rho_{\mu}^a  =  \overline{\rho}^a_{\mu}+r_{0}\delta_{\mu
0}\delta^{3a}$
\\ \hline
\end{tabular}
\end{center}
\vspace{0.5cm}
 In the field equations the
derivative terms are neglected and only time-like components of
the vector mesons will survive if one assumes homogenous and
isotropic infinite matter. The field equations derived from the
Lagrange function at the mean field level are
\begin{equation}
m_{\sigma}^2s_0+g_3s_0^2+g_4s_0^2=\sum_Bg_{\sigma
B}m^2_{eff,B}S(m_{eff,B},k_{F,B})
\end{equation}
\begin{equation}
m_{\omega}^2w_{0}+ c_3w_{0}^3=\sum_Bg_{\omega B}n_B
\end{equation}
\begin{equation}
m_{\rho}^2r_0=\sum_Bg_{\rho B}I_{3B}n_B
\end{equation}
\begin{equation}
m_{\sigma*}^2s_0^*=\sum_Bg_{\sigma^*B}m^2_{eff,B}S(m_{eff,B},k_{F,B})
\end{equation}
\begin{equation}
m_{\phi}^2f_0=\sum_Bg_{\phi B}n_B.
\end{equation}
The function $S(m_{eff,B},k_{F,B})$ is expressed with the use of
the integral
\begin{equation}
S(m_{eff,B},k_{F,B})=\frac{2J_B+1}{2\pi^2}\int_0^{k_{F,B}}\frac{m_{eff,B}}{\sqrt{k^2+m_{eff,B}}}k^2dk
\end{equation}
where $J_B$ and $I_{3B}$ are the spin and isospin projections of
baryon $B$, $k_{F,B}$ is the Fermi momentum of species $B$,
$n_{B}=\gamma_Bk_{F_{B}}^3/6\pi^2$ ($\gamma_B$ stands for the
spin-isospin degeneracy factor which equals 4 for nucleons and
$\Xi$ hyperons and 2 for $\Lambda$).
\newline
 The obtained Dirac equation
for baryons has the following form
\begin{equation}
(i\gamma ^{\mu }\partial_{\mu }-m_{eff,B}-g_{\omega
B}\gamma^{0}\omega_{0}-g_{\phi B}\gamma^{0}f_{0})\psi_B =0
\end{equation}
with $m_{eff,B}$ being the effective baryon mass generated by the
baryon and scalar fields interaction and defined as
\begin{equation}
m_{B,eff}=m_B-(g_{\sigma B}s_0+g_{\sigma^*B}s_0^*).
\label{masseff}
\end{equation}
The total energy of the system is given by
\begin{eqnarray}
\varepsilon  &=& \frac{1}{2}m_{\rho}^2r_{0}^2+\frac{1}{2}m_{\phi
}f_0^2+\frac{1}{2}m_{\sigma^* }(s_0^*)^2 +\frac{1}{2}m_{\omega
}^{2}w_{0}^{2}\\ \nonumber &+&\frac{3}{4}c_3w_{0}^{4}
+U(s_0)+\epsilon _{B} \label{en:dens}
\end{eqnarray}
where $\varepsilon _{B}$ equals
\begin{equation}
\varepsilon_{B}=\sum_{B}\frac{1}{3\pi^2}\int_{0}^{k_{F,B}}k^2dk\sqrt{k^2+m_{eff,B}^2}.
\end{equation}
\section{Parameters.} The considered model does not include
$\Sigma$ hyperons due to the remaining uncertainty of the form of
their potential in nuclear matter at saturation density
\cite{lamsig}, \cite{ksi}, \cite{ksi2}. The parameters that enter
the Lagrangian function are collected in Tables \ref{tab:TM1} and
\ref{tab:sscalar}. They are the standard TM1 parameter set
\cite{tm1} supplemented by hyperon-meson coupling constants. In
the scalar sector the scalar coupling of the $\Lambda$ and $\Xi$
hyperons requires constraining in order to reproduce the estimated
values of the potential felt by a single $\Lambda$ and a single
$\Xi$ in saturated nuclear matter
\begin{eqnarray}
U^{(N)}_{\Lambda}(\rho_0)&=&g_{\sigma
\Lambda}s_0(\rho_0)-g_{\omega \Lambda}w_0(\rho_0)\simeq 27-28 MeV\\
\nonumber
 U^{(N)}_{\Xi}(\rho_0)&=&g_{\sigma
\Xi}s_0(\rho_0)-g_{\omega \Xi}w_0(\rho_0)\simeq 18-20 MeV.
\end{eqnarray}
Assuming the SU(6) symmetry for the vector coupling constants and
determining the scalar coupling constants from the
potential depths, the hyperon-meson couplings can be fixed.\\
 The  strength of hyperon  couplings
to strange meson $\sigma^{\ast}$ is restricted through the
following relation
\begin{equation}
U^{(\Xi)}_{\Xi}\approx U^{(\Xi)}_{\Lambda}\approx
2U^{(\Lambda)}_{\Xi}\approx 2U^{(\Lambda)}_{\Lambda}.
\label{eq:pots}
\end{equation}
which together with the estimated value of hyperon potential
depths in hyperon matter provides effective constraints  on scalar
coupling constants to the $\sigma^{\ast}$ meson. The currently
obtained value of the $U^{(\Lambda)}_{\Lambda}$ potential at the
level of 5 MeV permits the existence of additional parameter set
which reproduces this weaker $\Lambda\Lambda$ interaction. In the
text this parameter set is marked as weak, whereas strong denotes
the stronger $\Lambda\Lambda$ interaction for
$U^{(\Lambda)}_{\Lambda}\simeq 20$ MeV. The relation \ref{eq:pots}
indicates that for the weak $\Lambda\Lambda$ interaction the
potential $U^{(\Xi)}_{\Xi}$ takes the value $\simeq 10$ MeV
whereas for the strong interaction $U^{(\Xi)}_{\Xi}\simeq 40$ MeV.
\begin{table}
\begin{center}
\begin{tabular}{llllllll}
  \hline
&$m_{\sigma}$[MeV]&$g_{\sigma N}$&$g_{\omega N}$&$g_{\rho N}$&$c_3$&$g_3$&$g_4$\\
\hline TM1&511.198&10.0289&12.6139&4.6322&71.3075&7.2325&0.6183\\
\hline TMA&519.151&10.055&12.842&7.6&151.59&0.328& 38.862\\\hline
\end{tabular}
\caption{Chosen parameter sets.}\label{tab:TM1}
\end{center}
\end{table}
\begin{table}
\begin{center}
\begin{tabular}{ll|l|l|l|l}
  \hline
&&$g_{\sigma\Lambda}$&$g_{\sigma\Xi}$&$g_{\sigma^{\ast}\Lambda}$&$g_{\sigma^{\ast}\Xi}$\\
\hline TM1&weak&6.2380&3.1992&3.7257&11.5092\\ \cline{2-6} &strong&6.2380&3.1992&7.9429&12.5281\\
\hline TMA&weak&6.2421&3.2075&4.3276&11.7314\\\cline{2-6} &strong&6.2421&3.2075&8.2580&12.7335\\
\hline
\end{tabular}
\caption{Strange scalar sector parameters.}\label{tab:sscalar}
\end{center}
\end{table}
The vector coupling constants for hyperons are determined from
$SU(6)$ symmetry \cite{shen} as
\begin{eqnarray}
\frac{1}{2}g_{\omega \Lambda}=\frac{1}{2}g_{\omega
\Sigma}=g_{\omega \Xi}=\frac{1}{3}g_{\omega N}  \\ \nonumber
\frac{1}{2}g_{\rho \Sigma}=g_{\rho \Xi}=g_{\rho N}; g_{\rho
\Lambda}=0  \\ \nonumber 2g_{\phi \Lambda}=2g_{\phi
\Sigma}=g_{\phi \Xi}=\frac{2\sqrt{2}}{3}g_{\omega N}.
\end{eqnarray}
\section{Infinite symmetric strange hadronic matter.}
For the symmetric strange hadronic matter there  no isospin dependance,
and there is no contribution coming from the $\rho$ meson field. There
are only two conserved charges: the baryon number
$n_b=n_{\Lambda}+n_{N}+n_{\Xi}$
 and the strangeness number
$n_s=n_{\Lambda}+2n_{\Xi}$. These two conserved charges allow one
to define the parameter which specifies the strangeness contents
in the system and is strictly connected to the appearance of
particular hyperon species in the model
\begin{equation}
fs =\frac{n_s}{n_b}=\frac{n_{\Lambda}+2n_{\Xi}}{n_b}.
\end{equation}
In a multi-strange system, for sufficient number density of
$\Lambda$ hyperons, the process $\Lambda + \Lambda \rightarrow \Xi
+N$, where $N$ stands for nucleon, becomes energetically allowed.
Thus, beside $N$ and $\Lambda$ also $\Xi^{-}$ and $\Xi^{0}$
hyperons have to contribute to the composition of strange hadronic
matter. In general the chemical equilibrium conditions for the
processes $\Lambda +\Lambda \rightarrow n+\Xi^{0}$ and $\Lambda
+\Lambda \rightarrow p+\Xi^{-}$ are established by the following
relations between chemical potentials
\begin{equation}
2\mu_{\Lambda}=\mu_{n}+\mu_{\Xi^{0}} \hspace{0.7cm}
2\mu_{\Lambda}=\mu_{p}+\mu_{\Xi^{-}}.
\end{equation}
This relation for symmetric matter can be rewritten as
$2\mu_{\Lambda}=\mu_N+\mu_{\Xi}$. The chemical potential $\mu_{B}$
($B=N, \Lambda ,\Xi$) through the Hugenholtz-van-Hove theorem is
related to the Fermi energy of each baryon in the following way
\begin{equation}
\mu_{B}=\sqrt{m_{B,eff}^2+k_{F,B}^2}+g_{\omega B}w_{0}+g_{\phi
B}f_0. \label{potchem}
\end{equation}
The binding energy of the system can be obtained from the
following relation
\begin{equation}
Eb=\frac{1}{n_b}(\varepsilon
-Y_Nm_N-Y_{\Lambda}m_{\Lambda}-Y_{\Xi}m_{\Xi})
\end{equation}
where $Y_{B} (B=N,\Lambda ,\Xi$) denotes concentrations of
particular baryons.
\section{Results}
\begin{figure}
\subfigure {\includegraphics[  width=8.15cm]{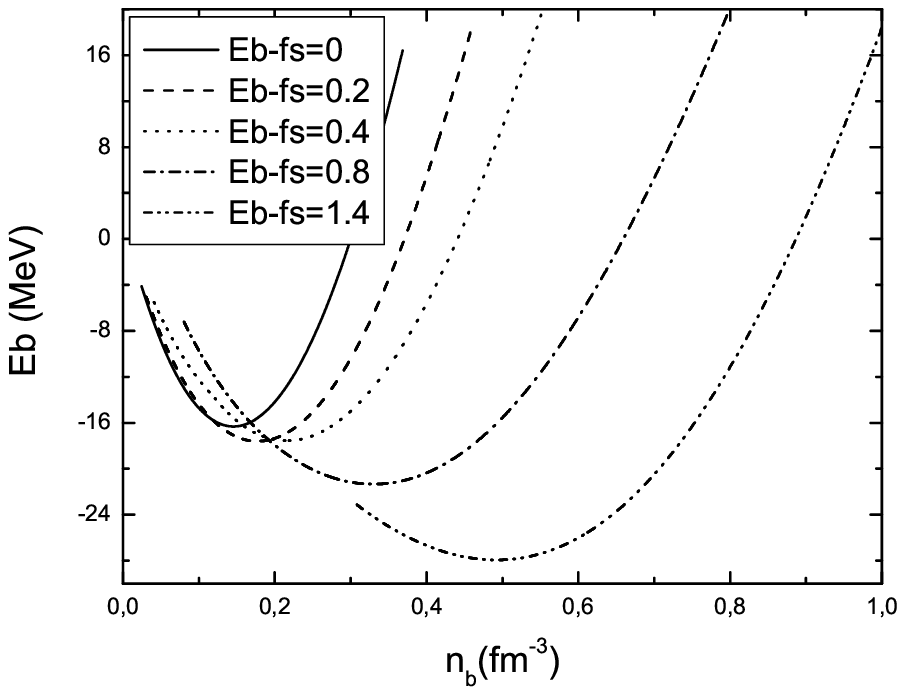}}
\subfigure {\includegraphics[  width=8.15cm]{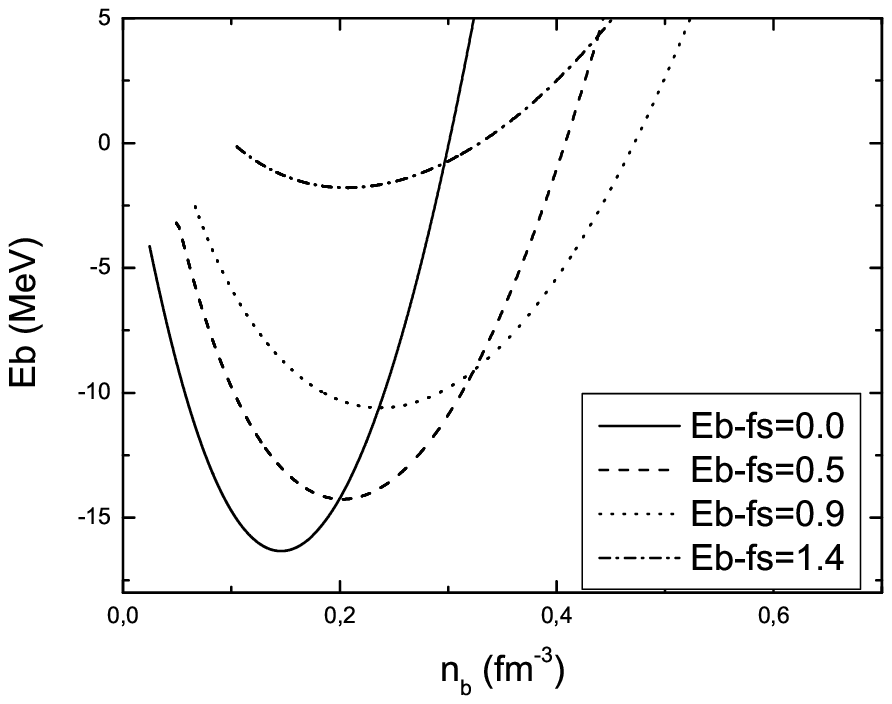}}
\caption{Density dependence of the binding energy $Eb$  for the
TM1 parameter set obtained for different values of strangeness
fraction $fs$. The left panel presents results for the strong
 and the right panel for the weak $Y-Y$
interaction models.} \label{Ebnb}
\end{figure}

\begin{figure}
\subfigure {\includegraphics[  width=8.15cm]{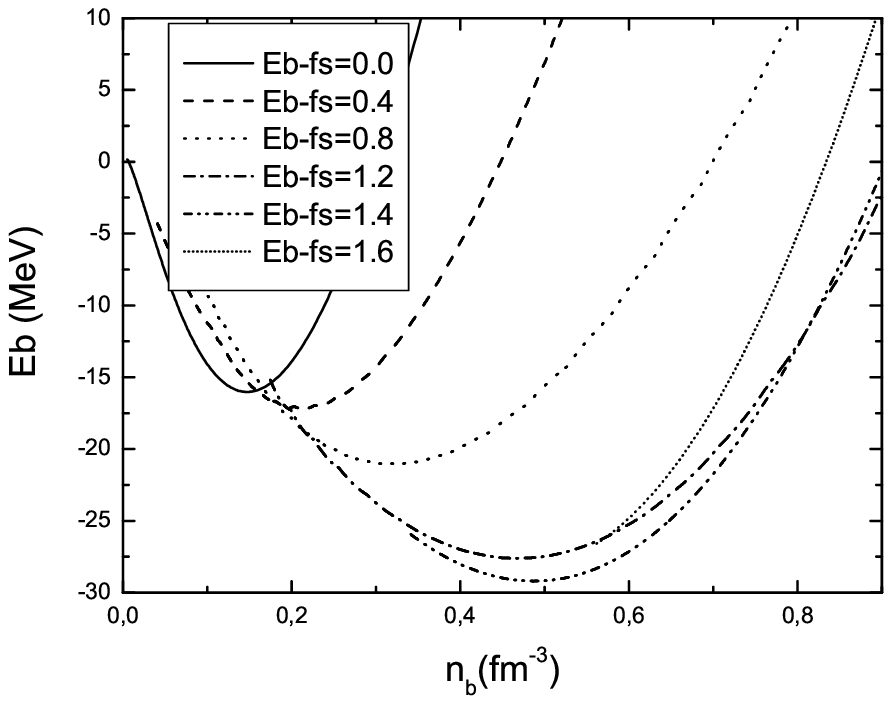}}
\subfigure {\includegraphics[  width=8.15cm]{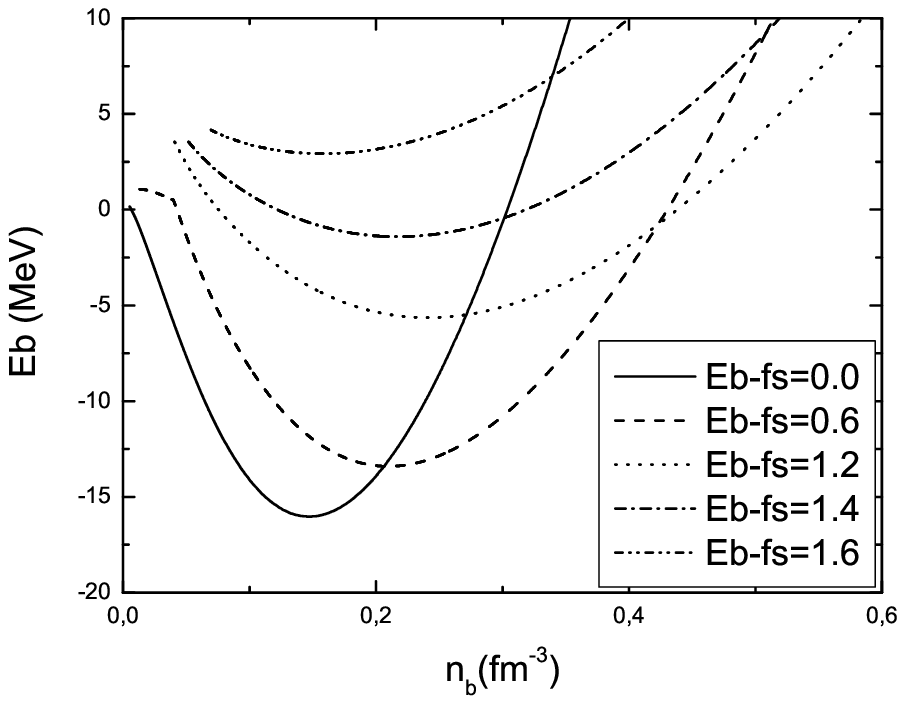}}
\caption{Density dependence of the binding energy $Eb$  for the
TMA parameter set obtained for different values of strangeness
contents $fs$. The left panel presents results for the strong
$Y-Y$ interaction, the right panel for the weak interaction
model.} \label{EbnbTMA}
\end{figure}

\begin{figure}
\subfigure {\includegraphics[  width=8.15cm]{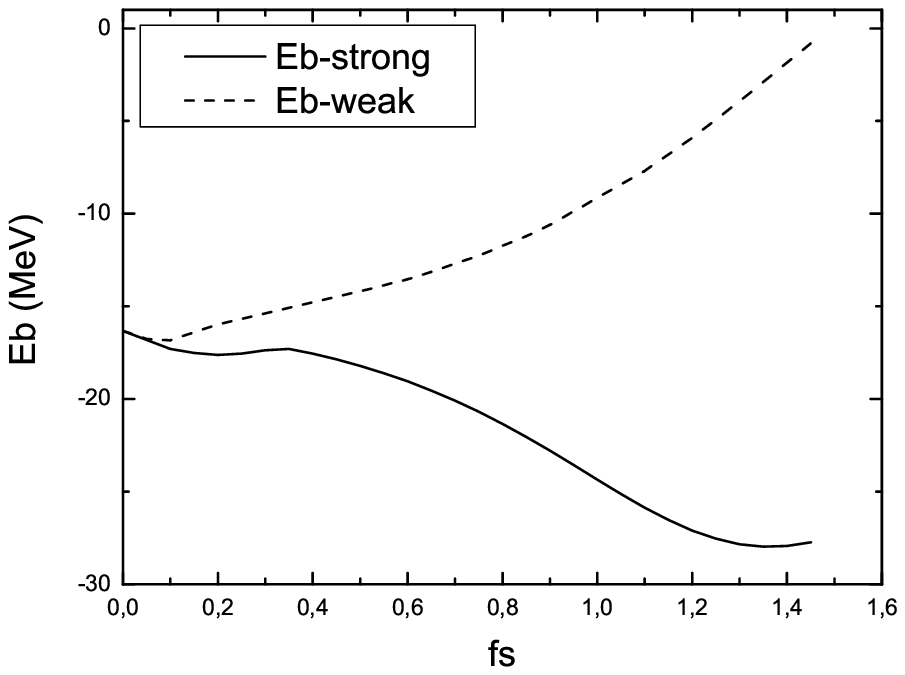}}
\subfigure {\includegraphics[  width=8.15cm]{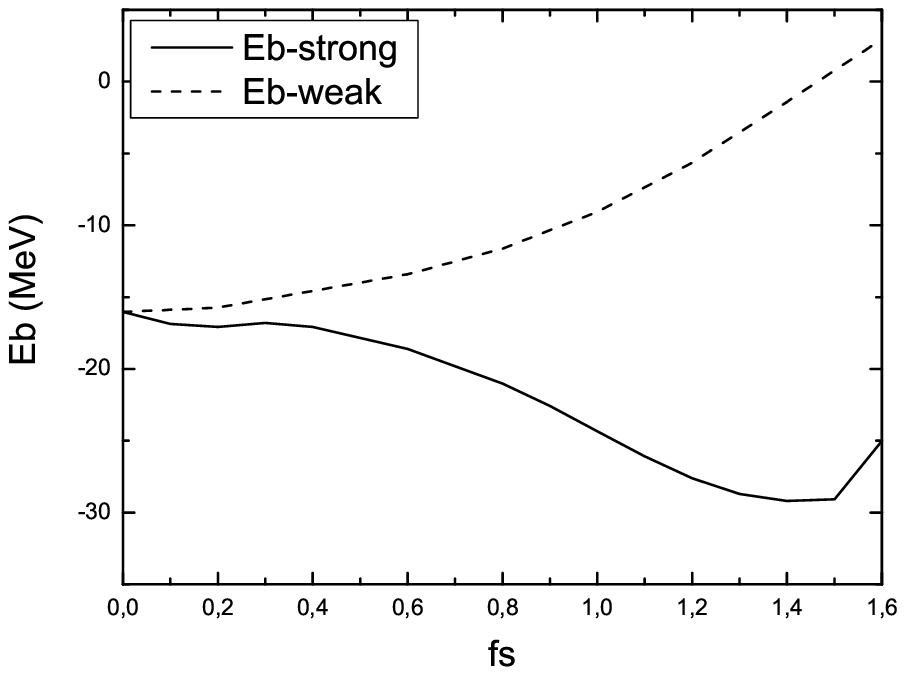}}
\caption{The minimized energy per baryon  for the TM1(left panel)
and TMA (right panel) parameter sets calculated for  the strong
and weak $Y-Y$ interaction models.} \label{ebfs}
\end{figure}

\begin{figure}
\centering
\includegraphics[width=10cm]{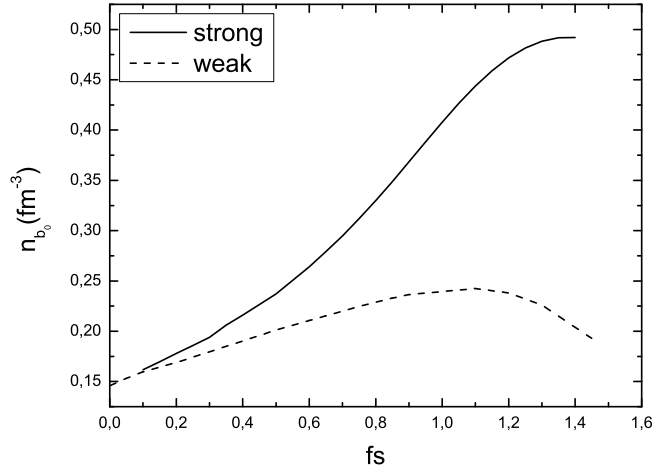}
\caption{The equilibrium density $n_{b_{0}}$ as a function of
strangeness contents $fs$.} \label{nb0fs}
\end{figure}

 The density dependance  of binding energies of multi-strange system involving
nucleons, $\Lambda$ and $\Xi$ hyperons  for different parameter
sets are presented in Fig.\ref{Ebnb} and Fig.\ref{EbnbTMA}. For
each parameterization two separated cases are considered, namely
the  strong and weak $Y-Y$ interactions. Individual lines
represent binding energies obtained for different values of
strangeness fraction $fs$. In all cases the value $fs=0$
corresponds to the state when only nucleons are present in the
system and the equation of state characteristic to nuclear matter
is reproduced.
The equilibrium density $n_{b_{0}}$ can be obtained by minimizing
the binding energy with respect to the baryon number density
$n_B$. Increasing the value of the parameter $f_s$, what is
equivalent with the increasing value of the strangeness contents
in the matter, the binding energy for each fixed value of the
parameter $fs$ has been calculated. The minimum values of  binding
energies for the fixed values of $fs$ have been determined and the
results are plotted in Fig.\ref{ebfs}.
For both parameterizations (TM1 and TMA), in the case of strong
$Y-Y$ interaction the increasing value of $fs$ leads to more bound
system with the minimum shifted towards higher densities. Contrary
to this situation the increasing value of strangeness contents in
hyperon matter characterized by weak $Y-Y$ interaction gives
shallower minima in the result. In Fig \ref{nb0fs} the equilibrium
density $n_{b_{0}}$ as a function of strangeness contents for the
two cases mentioned above is presented. The results are depicted for
the TM1 parametrization.
\begin{figure}
\subfigure {\includegraphics[  width=8.15cm]{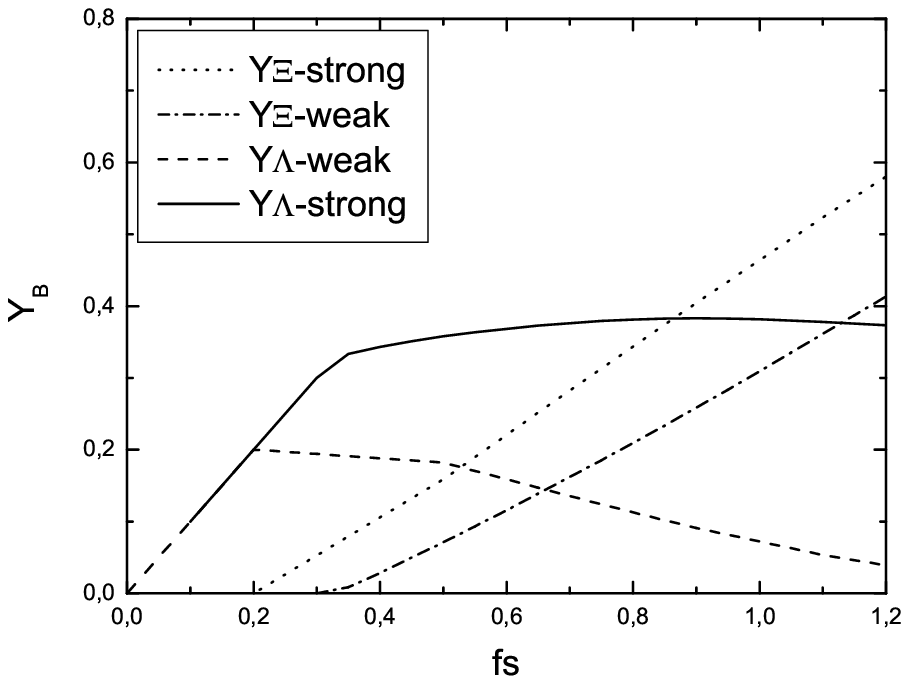}}
\subfigure {\includegraphics[  width=8.15cm]{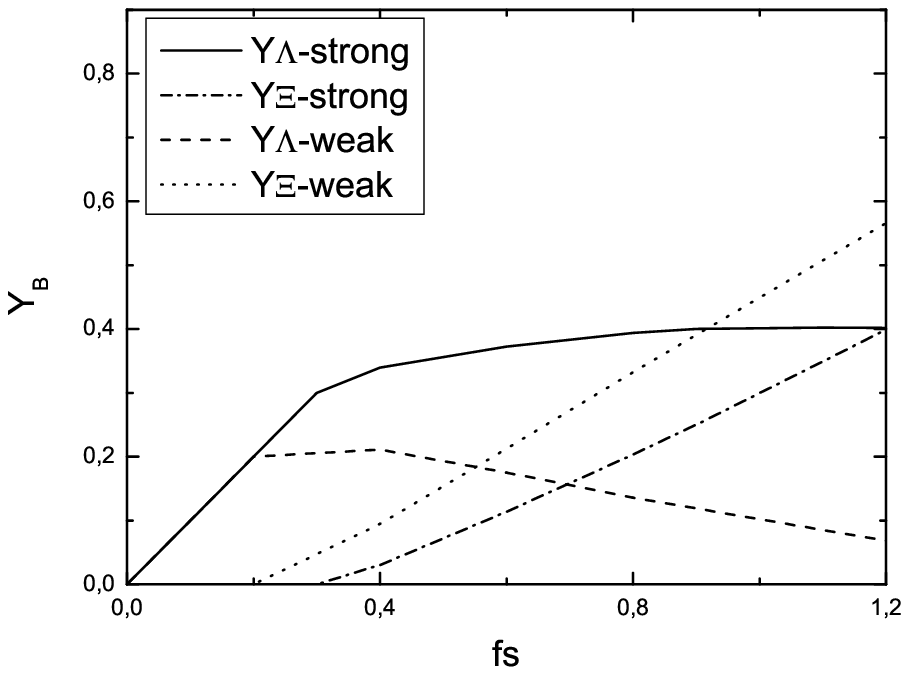}}
\caption{Relative concentrations of $\Lambda$ and $\Xi$ hyperons
in strange hyperon matter  for the TM1 (left panel) and TMA (right
panel) parameter sets.} \label{Yfs}
\end{figure}
Relative concentrations of $\Lambda$ and $\Xi$ hyperons in the
case of symmetric strange hadronic matter are presented in
Fig.\ref{Yfs}. In both cases the onset of $\Lambda$ hyperon is
followed by the onset of $\Xi$ hyperon. The populations of
$\Lambda$ hyperons obtained in the TM1 and TMA-weak models  are
reduced in comparison with those calculated for the strong $Y-Y$
interaction models. Concentrations of $\Xi$ hyperons are higher
for the weak TM1 and TMA models. In the case of weak $Y-Y$
interaction model the $\Xi$ hyperon thresholds are shifted towards
lower strangeness fractions. This has an influence on the
properties of neutron star matter.
\section{Hyperon star matter.}
\begin{figure}
\subfigure {\includegraphics[  width=8.15cm]{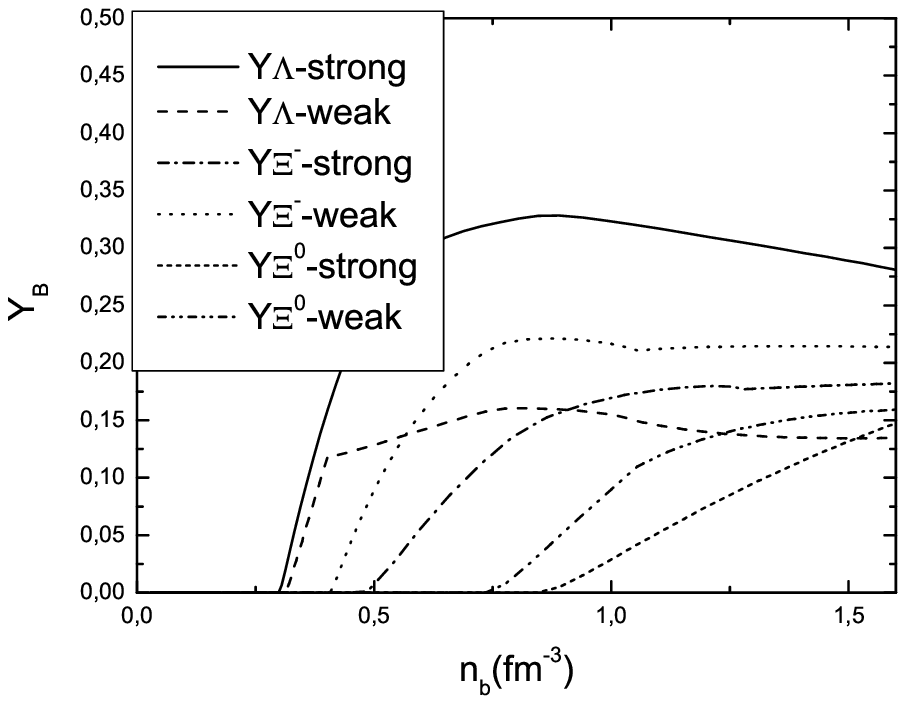}}
\subfigure {\includegraphics[  width=8.15cm]{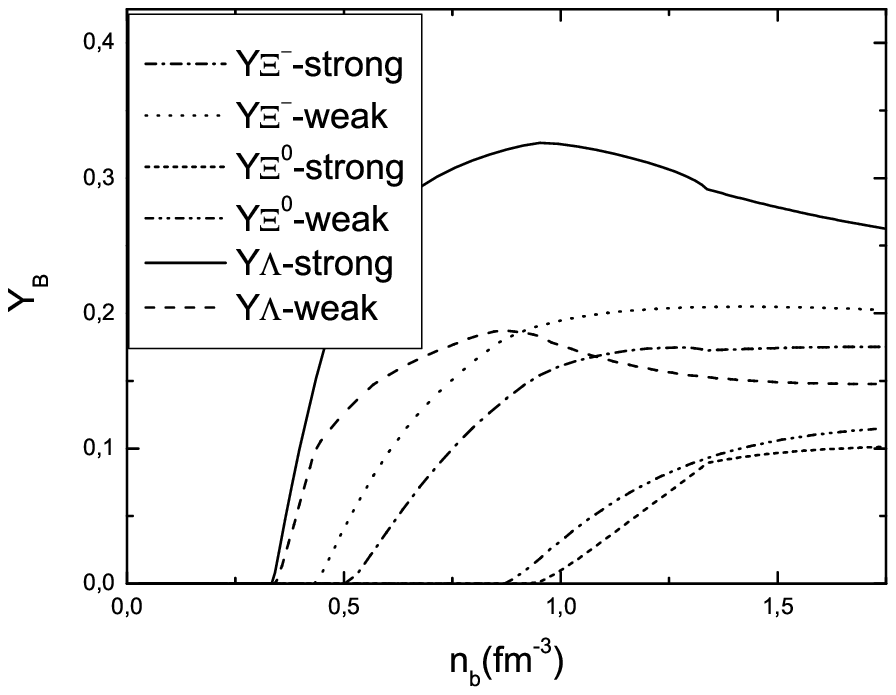}}
\caption{Relative concentrations of hyperons in hyperon star
matter as a function of baryon number density for the TM1 model
(left panel) and TMA model (right panel).} \label{YsnBstar}
\end{figure}

\begin{figure}
\subfigure {\includegraphics[  width=8.15cm]{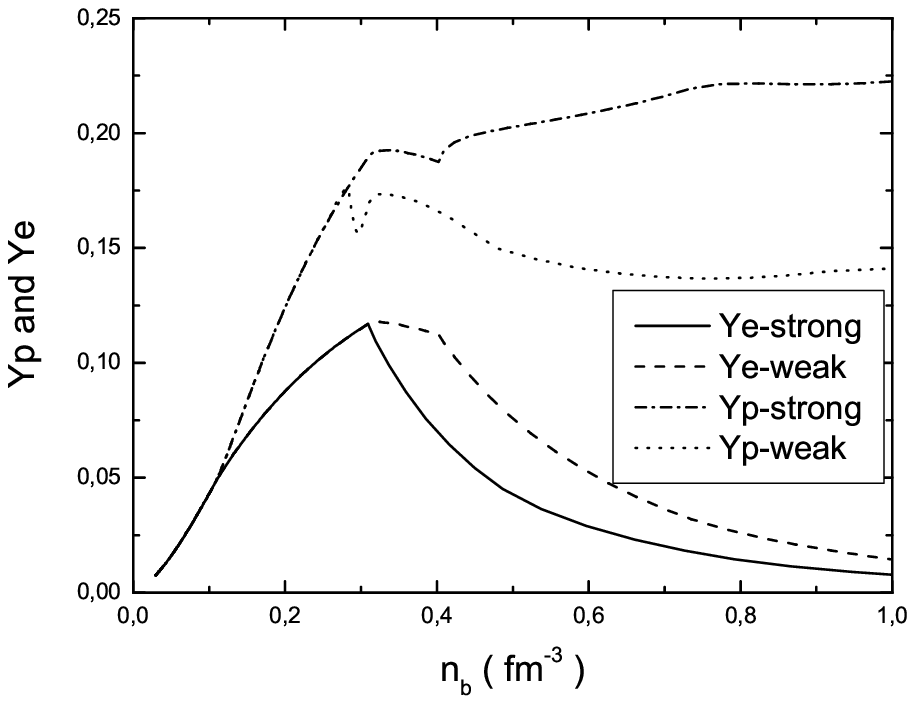}}
\subfigure {\includegraphics[  width=8.15cm]{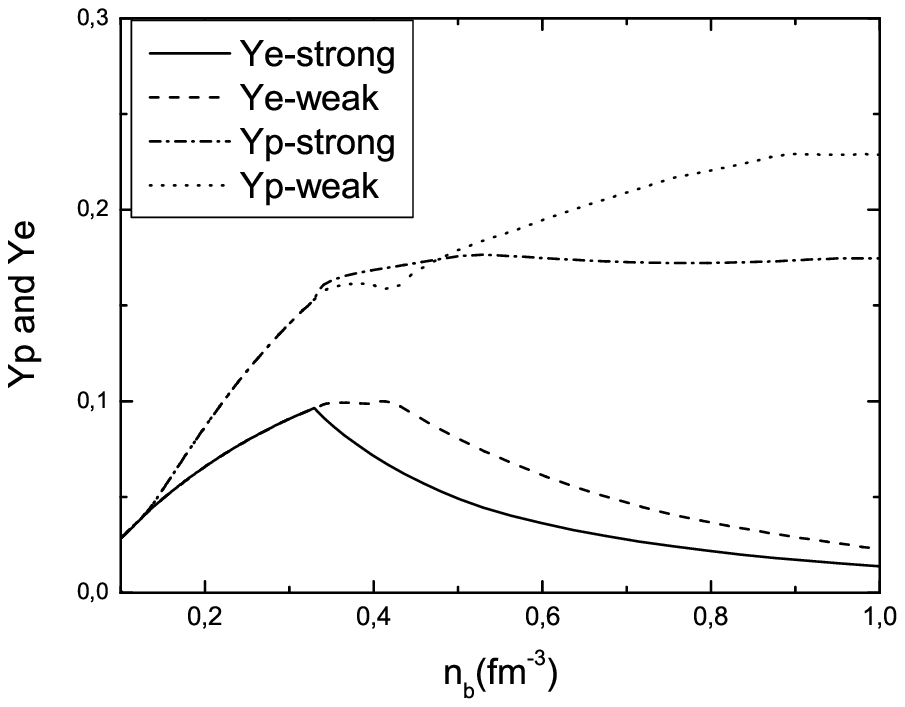}}
\caption{Concentrations of protons and electrons in hyperon star
matter as a function of baryon number density for the TM1 model
(left panel) and TMA model (right panel).} \label{YlpnBstar}
\end{figure}

\begin{figure}
\subfigure {\includegraphics[  width=8.15cm]{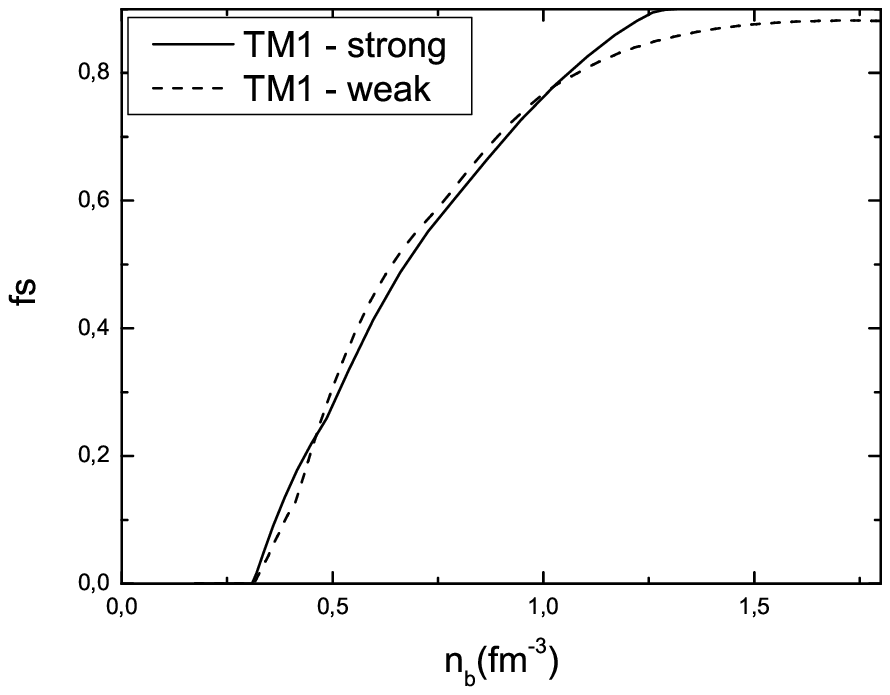}}
\subfigure {\includegraphics[  width=8.15cm]{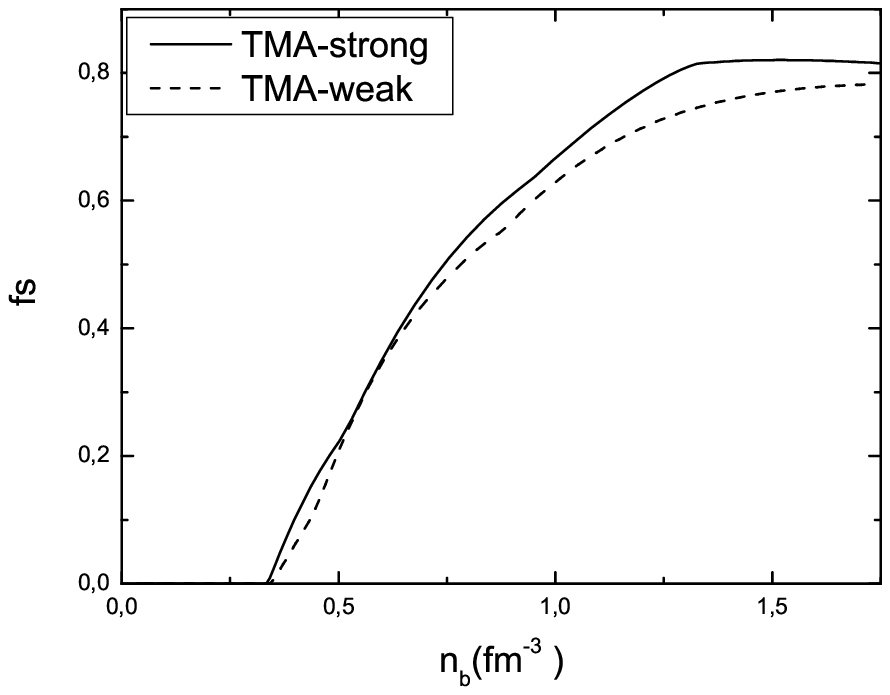}}
\caption{The strangeness contents as a function of baryon number
density for the TM1 model (left panel) and TMA model (right
panel).} \label{fsnbstar}
\end{figure}

\begin{figure}
\subfigure {\includegraphics[  width=8.15cm]{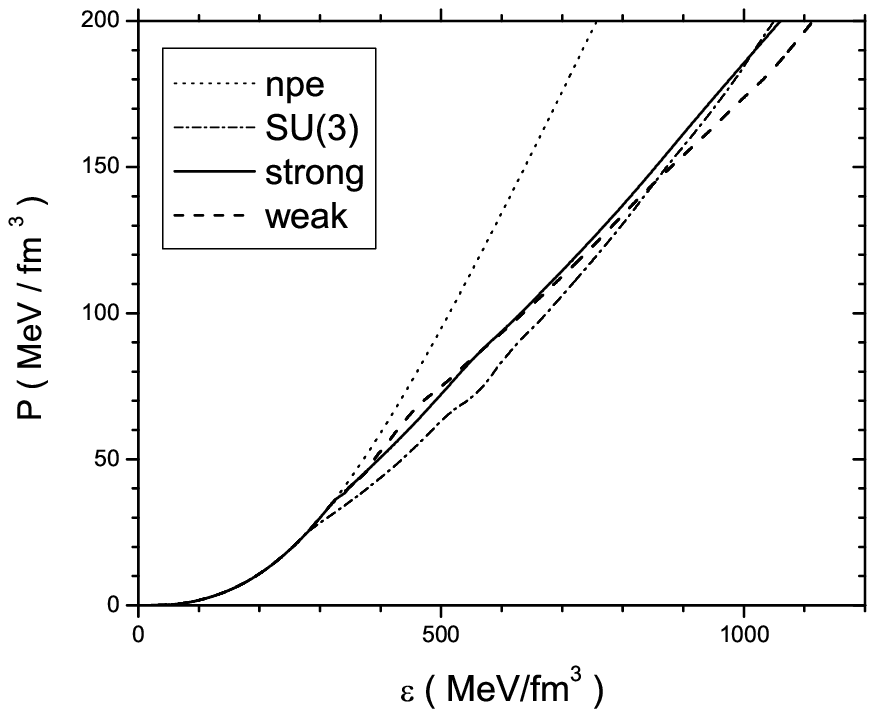}}
\subfigure {\includegraphics[  width=8.15cm]{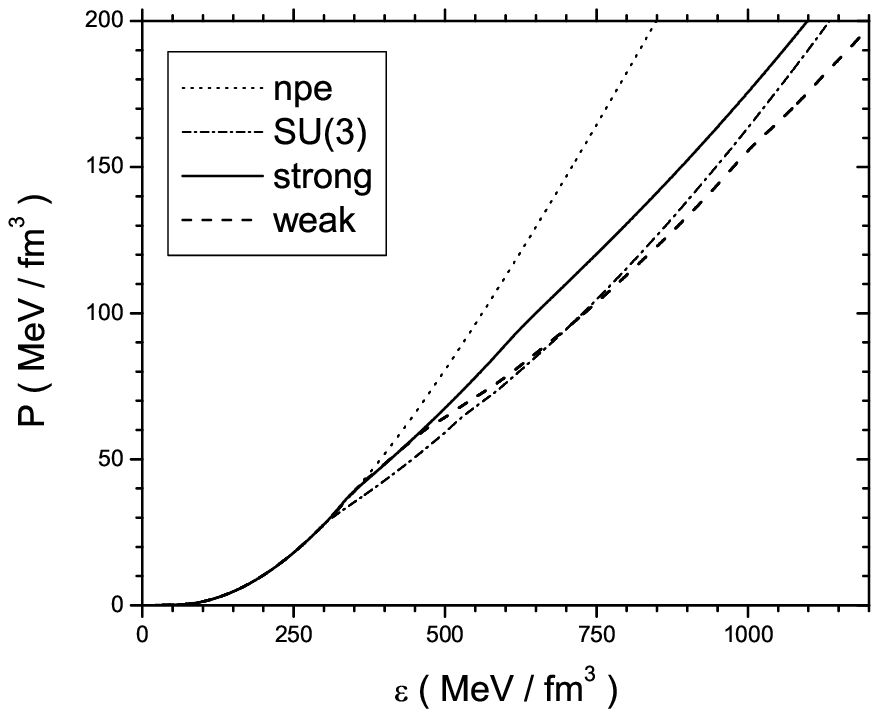}}
\caption{Equation of state for the TM1 (left panel) and TMA models
(right panel).} \label{fig:eos}
\end{figure}

\begin{figure}
 \subfigure {\includegraphics[width=8.15cm]{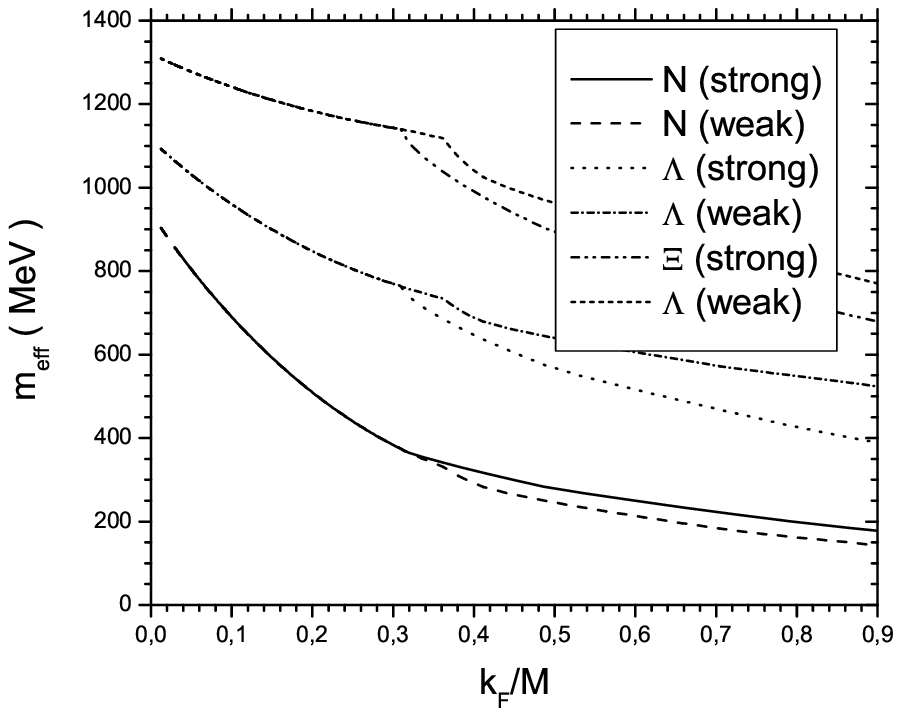}}
 \subfigure {\includegraphics[width=8.15cm]{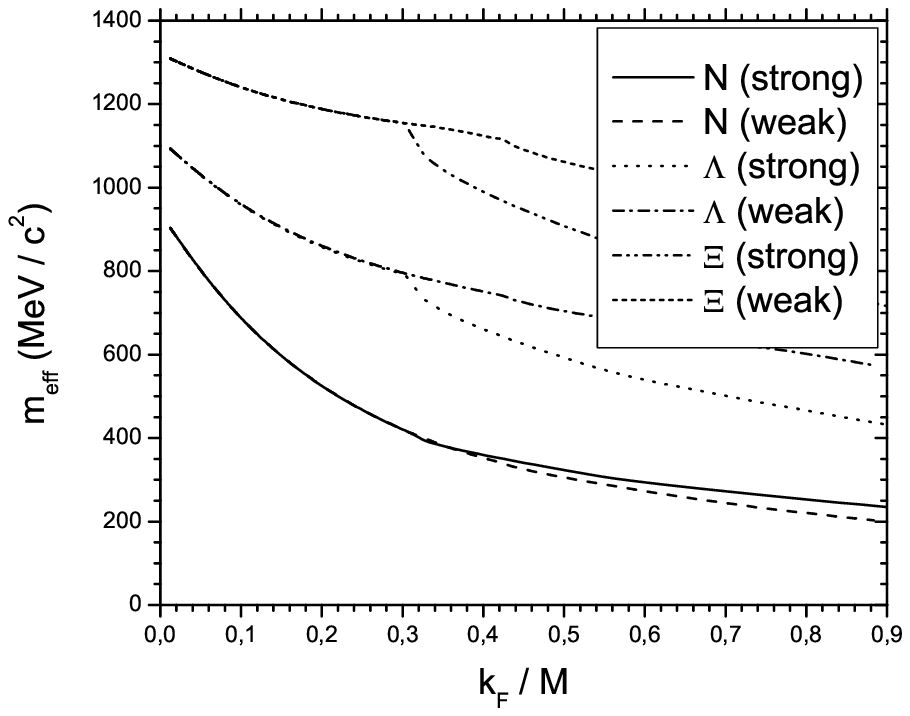}}
 \caption{Density dependance of the
effective baryon masses  for the TM1 (left panel) and TMA models
(right panel).} \label{fig:masseff}
\end{figure}

\begin{figure}
\subfigure {\includegraphics[  width=8.15cm]{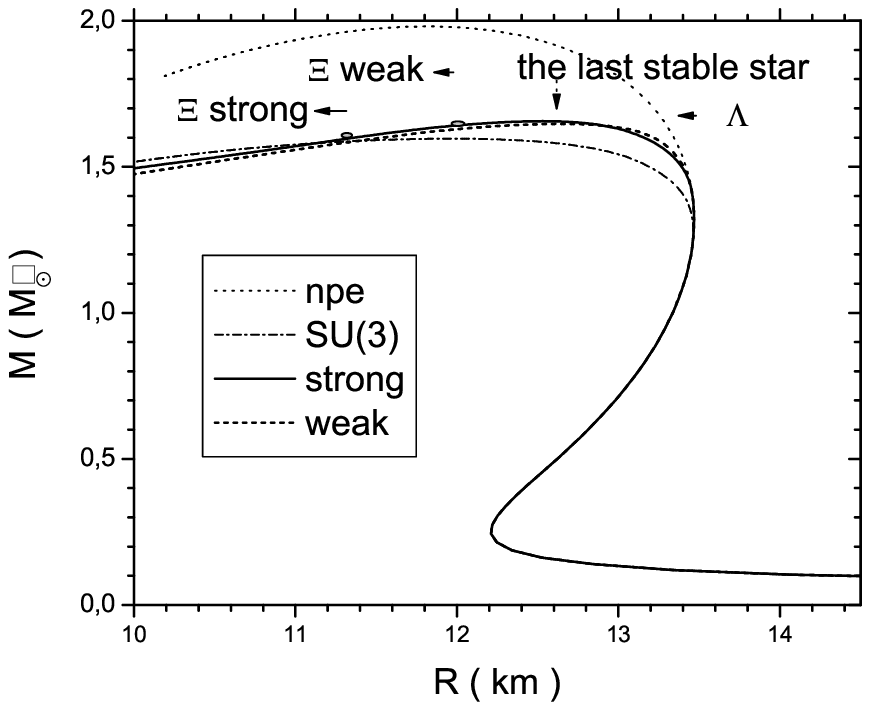}}
\subfigure{\includegraphics[  width=8.15cm]{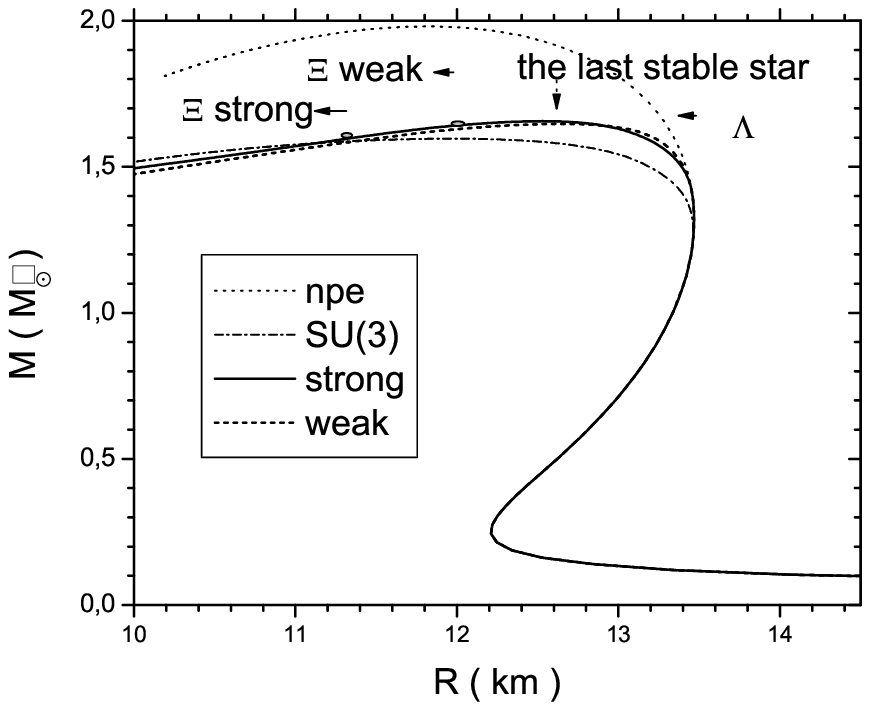}}
\caption{The mass radius relation for the TM1 (left panel) and TMA models (right
panel). Dots represent the configurations in which $\Xi$ hyperons
appear.} \label{fig:rm}
\end{figure}

\begin{figure}
\centering
\includegraphics[width=10cm]{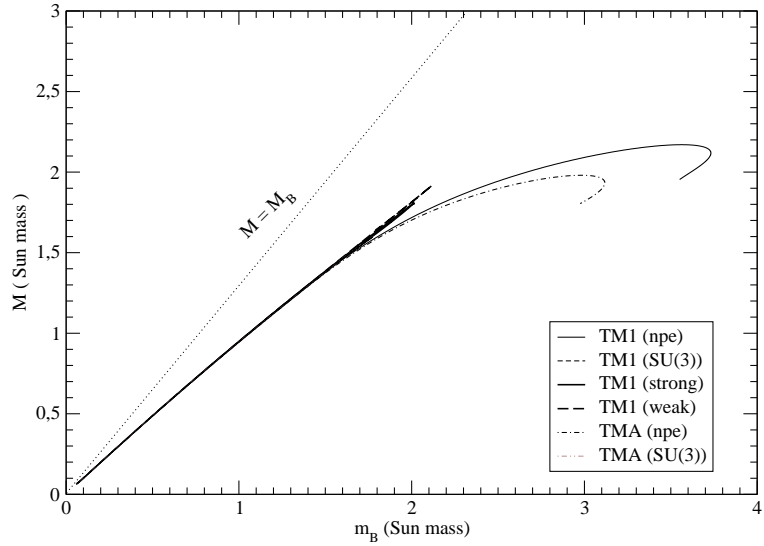}
\caption{The mass - baryon mass relation.} \label{fig:mb}
\end{figure}

Neutron star interiors and  relativistic heavy ion collisions
offer suitable environment for the existence of multi-strange
hyperon system. In the case of relativistic heavy ion collisions
hot and dense nuclear matter is probed whereas neutron star
interiors represent the density dominated scenario ($T \sim 0$).
 The analysis of the role of strangeness in nuclear
structure in the aspect of multi-strange system is of great
importance for neutron star matter. An imperfect knowledge of the
neutron star matter equation of state, especially in the presence
of hyperons, causes many uncertainties in determining neutron star
structure. It seems to be very important to estimate the influence
of nucleon-hyperon and hyperon-hyperon interaction on the equation
of state. The comparison between  weak interaction time scales
($10^{-10}$ s) and a time scale connected with the lifetime of a
relevant star  indicates that there is a difference between the
neutron star matter constrained by charge neutrality and
generalized $\beta$ equilibrium  and the matter in high energy
collisions; the latter matter is constrained by isospin symmetry
and strangeness conservation. The  condition of $\beta$
equilibrium in the case of neutron star matter implies the
presence of leptons and is realized by adding electrons and muons
to the baryonic matter. The Lagrangian of free leptons has the
following form (\ref{lg:lep})
\begin{equation}
L_{l}=\sum_{L=e,\mu}\overline{\psi}_{L}(i\gamma ^{\mu
}\partial_{\mu }-m_{L})\psi_{L}. \label{lg:lep}
\end{equation}
In general, in neutron star matter  muons start to appear after
$\mu_e$ has reached the value equal to the muon mass. The
appearance of muons not only reduces the number of  electrons but
also affects  the proton fraction. The requirements of charge
neutrality
\begin{equation}
n_p=n_e+n_{\mu}+n_{\Xi^-}
\end{equation}
and equilibrium under the weak processes
\begin{equation}
B_1\rightarrow B_2+L \hspace{2cm}B_2+L\rightarrow B_1
\end{equation}
($B_1$ and $B_2$ denote baryons) is decisive in determining the
composition of the hyperon star matter. The equilibrium conditions
between baryonic and leptonic species, which are present in
hyperon star matter, lead to the following relations between their
chemical potentials
\begin{eqnarray}
\mu_p=\mu_n-\mu_e
\hspace{10mm}\mu_{\Lambda}=\mu_{\Xi^0}=\mu_n  \\
\nonumber \mu_{\Xi^-}=\mu_{n}+\mu_{e} \hspace{10mm}
\mu_{\mu}=\mu_{e}.
\end{eqnarray}
The relations which determine the chemical potentials
(\ref{potchem}) and effective baryon masses (\ref{masseff})
indicate that the composition of hyperon star matter is altered
when the strength of the hyperon-hyperon interaction is changed.
Fig.\ref{YsnBstar} presents fractions of particular strange baryon
species $Y_B$ as a function of baryon number density $n_B$ for TM1
and TMA parameterizations. Starting the analysis of these graphs
from moderate densities it is evident that $\Lambda$ hyperons are
the most abundant strange baryons. $\Lambda$ is also the first
strange baryon that emerges in hyperon star matter, it is followed
by $\Xi^-$ and $\Xi^0$ hyperons. For the weak and strong $Y-Y$
interaction models  the sequence of appearance of hyperons is the
same however, in the case of the weak  $Y-Y$ interaction the
threshold for the $\Xi^{-}$ and $\Xi^{0}$ hyperons are shifted
towards lower densities. Due to the repulsive potential of
$\Sigma$ hyperons their onset points are possible at very high
densities which are not relevant for neutron stars.
In the case of weak $Y-Y$ interaction  the population of $\Lambda$
hyperons is reduced
whereas relative concentrations of $\Xi^-$ and $\Xi^0$ hyperons are enhanced.\\
The populations of protons and leptons are also altered by the
change of the strength of hyperon coupling constants. Through the
requirement of charge neutrality and $\beta$-equilibrium condition
the onset points and concentrations of hyperons affect  the
negatively charged lepton and proton abundance. Populations of
protons for TM1 and TMA parameterizations increase for the weak
$Y-Y$ interaction models. Results are presented in
Fig.\ref{YlpnBstar}.
The hyperonization of matter in the presented models can be also
analyzed through the density dependence of the strangeness
fraction.
 The density dependance of the $fs$ parameter in the hyperon star matter
 is depicted in Fig.\ref{fsnbstar}.
For the TMA parameter set more hyperon rich matter is obtained for
the strong $Y-Y$ interaction model, the TM1 parameter set gives
the same result for very high densities, for moderate densities
the strangeness contents is higher for the  weak model.\\ The
equation of state of the system can be calculated from the
stress-energy density tensor $T_{\mu\nu}$ which is defined as
\begin{equation}
T_{\mu\nu}=\sum_{a}\partial_{\nu}\Phi^a(x)\frac{\partial
L(x)}{\partial (\partial^{\mu}\Phi^a(x))}-g_{\mu\nu}L(x)
\end{equation}
where $\Phi^a=(\phi_M,\psi_B,\psi_L)$ and
$\phi_M=(\sigma,\omega,\rho,\sigma^{\ast},\phi)$.
 The equation of state also has been constructed for the strong and weak $Y-Y$ interactions, for the
chosen TM1 and TMA parameterizations.  The results are presented
in Fig.\ref{fig:eos} where besides the equations of state obtained
for the parameterizations discussed in this paper two other
selected equations of state are shown. The first represents the
neutron star model (with zero strangeness) and the second the
hyperon star model with hyperon couplings derived from the quark
model \cite{ma}.
Generally the equation of state obtained for the weak $Y-Y$
interaction model is less stiff that for the strong model.
However, there is an energy density range for which the weak model
gives the stiffer equation of state than the strong $Y-Y$
interaction model. The obtained form of the equation of state is
influenced by considerably altered effective baryon
masses. The results are presented in Fig. \ref{fig:masseff}.
On specifying the equation of state the properties and structure
of neutron stars can be obtained from hydrostatic equilibrium
equations of Tolman, Oppenheimer and Volkoff. The application of
the composite equation of state constructed by adding Bonn and
Negele-Vauterin equations of state  allows us to calculate the
neutron star structure for the entire neutron star density span.
At higher densities the equation of state  depends on the nature
of strong interactions hence the strength of hyperon-hyperon
interactions should exert influence on neutron star parameters.
The calculated mass-radius relations for the determined forms of
the equations of state are presented in Fig.\ref{fig:rm}.
In Table \ref{tab:maxmas} the parameters of the maximum mass
configurations are collected.
\begin{table}
\begin{center}
\begin{tabular}{ll|l|l|l}
  \hline
&&$\rho_c [g/cm^{3}]$&R [km]& $M_{max}[M_{\odot}]$\\
\hline TM1&weak&6.16$\times 10^{14}$&12.88&1.82\\ \cline{2-5} &strong&6.60$\times 10^{14}$&12.65&1.80\\
\hline TMA&weak&6.04$\times 10^{14}$&12.64&1.64\\\cline{2-5} &strong&6.28$\times 10^{14}$&12.53&1.66\\
\hline
\end{tabular}
\caption{Neutron star parameters for the maximum mass
configurations.}\label{tab:maxmas}
\end{center}
\end{table}
From Fig.\ref{YsnBstar} it is evident that  $\Xi^{-}$ and $\Xi^0$
hyperons emerge at very high densities. Although their onset
points in the case of weak $Y-Y$ interaction models are shifted
towards lower densities, still their presence in stable neutron
star configuration is uncertain. In Table \ref{tab:xi} the  star
parameters for the appearance of $\Xi$ hyperons are collected.
\begin{table}
\begin{center}
\begin{tabular}{ll|l|l|l}
  \hline
&&$\rho_x [g/cm^{3}]$&R [km]& $M [M_{\odot}]$\\
\hline TM1&weak&7.62$\times 10^{14}$&12.55&1.81\\ \cline{2-5} &strong&9.40$\times 10^{14}$&11.99&1.78\\
\hline TMA&weak&8.08$\times 10^{14}$&12.07&1.63\\\cline{2-5} &strong&9.79$\times 10^{14}$&11.67&1.62\\
\hline
\end{tabular}
\caption{Neutron star parameters for the appearance of $\Xi$
hyperons.}\label{tab:xi}
\end{center}
\end{table}
The value of densities, masses and radii collected in the Table
\ref{tab:xi} indicate that $\Xi$ hyperons do not appear in stable
hyperon star  configurations for both parameter sets. Thus only
$\Lambda$ hyperon will be present in the composition of hyperon
star matter for the considered models. Fig.\ref{fig:rm} shows the
position of the configurations, which parameters are presented in
Table \ref{tab:xi}, on the mass-radius relations.\\
Neutron stars are purely gravitationally bound compact stars. The
gravitational binding energy of a relativistic star is defined as
a difference between its gravitational and baryon mass.
\begin{equation}
Eb_{g}=(M_{p}-m(R))c^{2}\end{equation} where \begin{equation}
M_{p}=4\pi \int
_{0}^{R}drr^{2}(1-\frac{2Gm(r)}{c^{2}r})^{-\frac{1}{2}}\rho
(r)\end{equation} Of considerable relevance is the numerical
solution of the above equation for the selected equations of
state.
 Fig. \ref{fig:mb} depicts the gravitational mass which includes
interactions versus the baryonic mass for all the considered
models.
\section{Conclusions.}
The main goal of this paper was to study the influence of the
strength of hyperon-hyperon interactions on the properties of the
hyperon star matter and on a hyperon star structure. It has been
shown that replacing the strong $Y-Y$ interaction model by the
weak one introduces large differences in the composition of a
hyperon star matter both in the strange and non-strange sectors.
There is a considerable reduction of $\Lambda$ hyperon
concentration whereas the concentrations of $\Xi^{-}$ and
$\Xi^{0}$ hyperons are enhanced. In the non-strange sector the
populations of protons and electrons are changed. The weak model
permits larger fractions of protons and electrons. The presence of
hyperons in general  leads to the softening of the equation of
state. For the employed weak $Y-Y$ interaction model there is a
density range for which the obtained equation of state is stiffer
than the one calculated with the use of the strong model. This is
clearly visible in the case of TM1 parameter set. For higher
densities the  weak model gives less stiff equation of state. The
behavior of the equation of state is directly connected with the
value of the maximum star mass. Equilibrium conditions namely
charge neutrality and $\beta$-equilibrium determine the
composition of the star. For both parameterizations the onset
points for $\Xi^-$ and $\Xi^0$ hyperons are localized at high
densities which are relevant to unstable branches of the
mass-radius relations. Thus the obtained stable hyperon star model
is composed of neutrons, protons, $\Lambda$ hyperons and leptons.
The appearance of $\Xi^-$ and $\Xi^0$ hyperons in neutron star
interior will be possible in a very special configuration. A
protoneutron star model with very high central density can lead to
a hyperon star with $\Xi$ and $\Lambda$ hyperons in its interior.
The possibility of the existence of such protoneutron star model
will be the subject of future investigations. One can compare the
obtained results with those presented in the paper by Schaffner et
al. \cite{SB}, where the analysis of the existence of the  third
family of stable compact stars has been performed for highly
attractive hyperon-hyperon interaction. However, recent
experimental data indicate for much weaker strength of $Y-Y$
interaction. Employing these data and the estimated value of the
$\Lambda$ well depth $U_{\Lambda\Lambda}\simeq 5$ MeV the result
of the paper \cite{SB} can not be confirmed.


\end{document}